\documentclass[12pt,english]{article}
\usepackage[T1]{fontenc}
\usepackage[latin1]{inputenc}
\usepackage{floatflt}
\usepackage{graphicx}
\usepackage{amssymb}

\makeatletter

 \newenvironment{lyxcode}
   {\begin{list}{}{
     \setlength{\rightmargin}{\leftmargin}
     \setlength{\listparindent}{0pt}
     \raggedright
     \setlength{\itemsep}{0pt}
     \setlength{\parsep}{0pt}
     \normalfont\ttfamily}%
    \item[]}
   {\end{list}}

\usepackage{a4}

\usepackage{babel}
\makeatother
\begin{document}

\title{Is a Rich {}``Vacuum'' Structure Responsible for Fermion and Weak-Boson
Masses}

\maketitle
\begin{lyxcode}
\begin{center}\textrm{\textsc{Fritz~W.~Bopp}}\\
\textrm{\textsc{University~of~Siegen}}\end{center}
\end{lyxcode}
\begin{abstract}
An unconventional \emph{{}``Cosmographical model''} for a generation
of fermion and week boson masses without electro-weak Higgs bosons
is outlined. It is based on a rich, non perturbative vacuum structure
taken to be an object of qualitative phenomenology. Numerous far reaching
and astonishing consequences are discussed.
\end{abstract}

\section{Introduction}

The standard model was developed a long time ago and it might be argued
that the field is closed as far as basic concepts at electroweak energies
are concerned. However, one should keep in mind these concepts have
actually never been tested in its central part~\cite{Higgs}. As
LHC starts taking data it is widely~\cite{Grinbaum} felt as appropriate
to recheck the basic premises in a sceptical way. There are four basic
points where the usual argumentation is not accepted. 

The first point concerns the \textbf{\emph{hierarchy problem}} in
particle physics. The non-vanishing Higgs field is part of the {}``vacuum''
structure. The understanding of the {}``vacuum'' cannot be separated
from cosmological considerations. Cosmology contains a massive - $60$
orders in magnitude \cite{Bousso07} - hierarchy problem in its {}``vacuum''
structure: The cosmological constant is usually taken to correspond
to the {}``vacuum'' energy density. The flatness of the universe
requires on the one hand a non-vanishing but tiny cosmological constant.
On the other hand the properties of a condensate have to reflect the
Grand Unification scale of the interactions when it was formed. A
solution to the puzzle has to be found in cosmology and an evolving
vacuum state~\cite{Neupane} makes most sense. 

For particle physics this means: Two scales, $M_{Planck}$ and $m_{dark\, energy}$,
are available. There is no need for a bridge between present particle
masses and the unification scale as the masses might just as well
be obtained from a present dark energy scale. Without the usual hierarchy
argument there is no need to stabilize Higgs masses invoking supersymmetry
or to shrink the gap to the Planck mass with extra dimensions.

The second point concerns the \textbf{\textit{universality of the
{}``vacuum''.}} To explain the symmetry breaking with four component
Higgs fields one writes down a potential for the neutral components
of these fields ( i.e. for $\phi\in\mathbb{C}$) \[
U(\phi,t)=bT(t)\:|\phi|^{2}+\Lambda(t)+\frac{\lambda}{4}\:(|\phi|^{2}-|\phi_{0}|^{2})^{2}\]
where $T(t)$ is taken as its temperature and $\Lambda(t)$ as corresponding
cosmological constant. A non-vanishing vacuum expectation value $<\phi>\,\ne0$
results.

The time dependence is usually not included and the resulting vacuum
field is taken to be a universal field theoretical object without
coordinate dependence. However, even in the standard model this is
not really correct. The {}``vacuum\char`\"{} has a role as a reservoir.
Consider left-handed electrons in a purely QED {}``gedanken'' synchrotron.
Their anomalous momentum causes them to oscillate between their left-
and right-handed states, i.e. to oscillate between $Q_{U(1)}=-\frac{1}{2}$
and a $Q_{U(1)}=-1$ states in an apparent violation of the $U(1)$-charge
conservation. The same applies to the $SU$(2) quantum numbers. To
compensate the change the Higgs filling the {}``vacuum\char`\"{}
has to slightly rotate between its two neutral components. This is
no problem but it implies a time dependence and by relativity a space
dependent {}``vacuum\char`\"{}. 

The {}``vacuum'' as a time dependent, evolving reservoir is the
central dictum of our model. Except for a decoupling from the usual
interactions in the visible world the \char`\"{}vacuum\char`\"{} is
considered as an {}``underworld'' filled with real objects. Of course
nowadays the \char`\"{}vacuum\char`\"{} has to be rather uniform even
on a cosmic scale. 

The third point argues for a \textbf{\textit{fermionic component}}
in the vacuum. In scalar field theories it is difficult to have stable,
non trivial solutions. The condensate concept of the vacuum is taken
from condensed matter physics~\cite{Anderson}. Almost all condensates
realized in nature involve fermions and the residual Fermi repulsion
is an essential parameter of the understanding of the extent of the
condensate. 

The last point concerns actual mass generation. The irregularity of
the needed values leads to the postulate that \textbf{\emph{masses
are determined in a to a degree chaotic process}}%
\footnote{According to Hawking~\cite{Hawking}: {}``It is hard to believe,
that so many and so irregular parameters of the various mass matrices
can be determined from first principles''. To support this argument
numerous tries could be cited (one example is~\cite{fbopp}). The
early evolution of the vacuum can be described by individual scattering
processes and a chaotic description is applicable. %
}\textbf{\emph{.}} This postulate of essentially random values is not
new%
\footnote{The Multiverse concept~\cite{Weinberg,Wilczek} assumes that the
symmetry breakdown determining the masses is chosen randomly in the
early universe constrained by anthropogenic considerations. The widely
varying spectrum of masses has then somehow to be protected from mixing
on the way down to the observed-mass-matrices scale. In some super
symmetric theories special {}``messenger'' fields were introduced~\cite{Haber}. %
}. Here we assume that the random process ends in a tumbled down vacuum
involving today`s mass scales. 

These points suggest a novel view of fermion masses. In the standard
Higgs model all fermion mass parameters are transferred to coupling
constants:\[
m_{ij}=g{}_{ij}<h>\]
The new view is based on an modified attribution:\[
m_{ij}=g<h{}_{ij}>\]
 The gauge group couplings should be part of pure physics and unique.
The multitude of mass parameters is transferred to Cosmography, i.e.
a rich vacuum%
\footnote{To attribute the flavor parameters to a basically statistical {}``vacuum\char`\"{}
structure~\cite{bopp:cosmo} was also advocated in~\cite{Hall_Salem_Watari}
in a less simple minded model. To have a separate Higgs for each fermion
is named {}``Private Higgs''~\cite{Porto}. 

This separation between fundamental physics and \emph{Cosmography}
is central. To calculate the weight of a air balloon on your desk
from first principles is bound to fail, as the barometric pressure
determined by largely chaotic \emph{Geograph}y enters. %
} that like geographical objects is naturally expected to be messy.
It means that today's multitude of {}``physical'' constants cannot
be calculated from first principles. It has to be attributed to a
partially chaotic structure of a particular {}``cosmographic vacuum''
in our zone of the universe. In this way many aspects of the model
are not derivable and the model is not the optimist´s choice. The
aim can just be a phenomenological parton model kind of understanding.

The required rich {}``vacuum'' structure is not far fetched. Even
in the {}``standard model'' the {}``vacuum'' is not really simple.
It is known to contain a certain amount of chiral condensates of gluonic
and fermionic nature and particular combination of the neutral Higgs
fields with a complicated potential not respecting a separation of
U(1) and SU(2). 

In this paper we will outline a particular condensation model based
on these points. Section 2 details the evolution scenario. The particular
choice of {}``post-desert-physics'' advocated should be taken as
generic. Also the details of the tumbling-down-vacuum description
is not intended as a cosmological model. The intent of these considerations
taken in a specific frame work is just to make sure that there is
no intrinsic inconsistency. Our aim is to understand the structure
of the present {}``vacuum'' needed for the known particles. 

Section 3 describes how the fermions get their masses. There are novel
ideas about apparent flavor changes and CP violation. It contains
speculation about the rules that govern the evolution of the vacuum.
Section 4 turns to the electroweak vector-boson masses. They tell
us about an important aspect of the {}``vacuum'', which relates
the weak mixing angle, the visible baryon asymmetry, and visible CP
violation in a conceptual way. Possible experimental observations
and tests follows in section~5~%
\footnote{A more general description of the basic framework was presented in
an earlier paper \cite{bopp:cosmo}. %
}.

\section{The Tumbling Down Condensation Process}

Following a standard SO(10) scenario we assume that the gauge theory
breaks at a scale of about $10^{15}\,$GeV eventually to SU(3)$_{C}$$\times$SU(2)$_{L}$$\times$U(1)$\times$U(1)$_{X}$
in a suitable unspecified way%
\footnote{The Cosmographic model has its own mechanism of fermion mass generation.
The inconsistency between SU(5) predictions with proton decay experiments
can be ignored as a sufficiently awkward attribution of charged leptons
to generations can eliminate or drastically reduce decays into charged
leptons. The decays into neutral leptons are then sensitive to the
second lepto-quark boson mass ($\tau\approx M_{Y}^{4}/m_{p}^{5}$)
that is sufficiently uncertain. %
}. The gauge bosons mixing subgroups have to obtain masses. Also a
mechanism for the U(1)$_{X}$ gauge boson mass is required. How this
might proceed~\cite{Ellis79} is beyond the scope of the paper. It
is assumed that a mechanism for the GUT scale symmetry breaking can
be found that does not lead to a significant vacuum energy density.
There could be some vacuum structure that sufficiently tumbles down
or the \char`\"{}proximity'' of the GUT scale to Planck scale might
help.

Independent of the details there have to be effective scalar bosons
needed by the gauge boson to become massive. The flavor independent%
\footnote{In contrast to top-condensate models~\cite{Lindner,Bardeen} the
cosmographic model does not allow for a special flavor dependence
in the Lagrangian. A fermionic {}``vacuum\char`\"{} was recently
also discussed in \cite{Quimbay}. %
} coupling constant of these bosons to fermions is unknown. A way to
implement the model is to assume it is sufficiently large to allow
for bound states that eventually can turn into a condensate. These
GUT scales bound states are modeled following the GeV scale $\sigma$-particle
vacuum known from chiral symmetry breaking. 

With different constituents many different such states exist. Massless
fermion-anti-fermion Yoyo-like states require transitions between
left- and right-handed fermions like $\overline{d{}_{R}}\, d{}_{R}\leftrightarrow\,\overline{d{}_{L}}\, d_{L}$.
The gauge bosons needed are still available at GUT scale. Besides
these $f\,\bar{f}$ states there also can exist bound states of multiple
fermionic (p.e. $f_{L}$$f_{L}$$f_{L}$$f_{R}$$f_{R}$$f_{R}$)
or multiple anti-fermionic content%
\footnote{If there is no lepton number conservation\cite{Klapdor,Gironi}  a
purely two component bound state $\overline{\nu_{R}}\,\nu_{R}$ can
also exist. It generates contributions to the Majorana masses of right-handed
neutrinos with the mechanism described in section 3. With a extremely
large density this vacuum state might even be the cause of the U(1)$_{X}$-mass. %
}.

Like the decoupling of the electromagnetic cosmic background radiation
during freeze out on a eV scale these bound states will largely decouple
from most gauge bosons. In this way they form {}``pre-vacuum'' states.
As discussed in section 3 and 4 for the {}``vacuum'' such {}``pre-vacuum''
states will start to give masses to the other fermions and gauge bosons.
In turn the created mass allows for tighter and eventually denser
bound states. This process goes on until todays {}``vacuum'' is
reached. 

In spite of their now massive content these bound states can have
arbitrarily small masses. The coupling from the gauge potential is
assumed to be strong enough to lead to negative {}``condensation''
potential compensating these masses:\[
V_{\mathrm{{fermion\, bound\: to\: bosons}}}+\sum m_{\mathrm{{fermion}}}\approx0\]
The potential energy should be considerably larger for multi-fermion
states~\cite{Froggatt:2008hc}. If the equation holds for $f\,\bar{f}$
states multi-fermion states appear to lead to a negative energy density.
However, the energy stays positive%
\footnote{Contrary to arguments stated from the cosmological side~\cite{Bousso07}~$\rho_{\Lambda}=0$
is special point also in particle physics. Such a particle creation
is well-known in hadronic string phenomenology~\cite{Andersson}.%
}. Approaching a vanishing value more and new bound states will be
produced until the fermion repulsion gets high enough to compensate
the negative energy density. 

The tumbling-down idea is that the almost zero energy vacuum states
are not formed initially. Initially $\approx0$ means zero within
GUT scale uncertainty. Radiating with their dipol moment they can
condense and continuously loose energy and entropy, like an atom being
caught by a crystal. 

Besides the processes discussed above there is a geometrical aspect.
The mass of the relevant vector bosons limits their dipol distance
$<|\: r_{F}-r_{F}^{_{-}}|>$ to a GUT scale value. After being produced
correspondingly localized they can reduce their energy by spreading
$<|\: r_{F}+r_{F}^{_{-}}|>\to$large. 

This extending process selects states with suitable vacuum quantum
numbers. As the photon and the gluon stays massless only charge and
color neutral states can participate in this extension process. The
other bound states will change or stay in the visible world. During
this extending process the overlap with usual particles will get weaker
and weaker making the decoupling stronger and stronger and decoupling
is the defining property of the cosmographical {}``vacuum''%
\footnote{There are other models with such an evolution in the \char`\"{}vacuum\char`\"{}
energy \cite{Bousso07,Neupane}. %
}. 

The central assumption for the tumbling down is that there is no intrinsic
mass scale%
\footnote{Such a situation is called gap-less in solid state physics\cite{Volovik,Yukalov}%
}. Contrary to the usual situation there is no lower limit as the {}``vacuum''
states can always reduce their energy further by spreading out. Without
any available scale the energy in a cell decreases as $\partial\rho_{vac.}^{-1}/\partial(t)=\kappa$
where $\rho_{vac.}$ is the energy of a {}``vacuum'' cell and $\kappa$
a dimensionless decay constant. This yields a simple power law $\rho_{vac.}\propto\rho_{initial}/(t-t_{0})$
. It connects the ratio of the initial almost GUT scale vacuum energy
and the cosmological observed one to age of the universe seen in a
particle physics scale, i.e. $\tau_{universe}\cdot M_{X(GUT)}$ .
On a conceptual level this offers a possible solution of the hierarchy
problem in cosmology~%
\footnote{This dimensional arguments is meant as simplest choice. Dynamical
dark energy models can include the Planck mass and a model dependent
power can be obtained. Important is the power law with a negative
exponent.

For a dynamical dark energy model connecting the Planck and the dark
energy scale \cite{Klinkhamer:2008nr} it is claimed on general grounds
that the smallness of the dark energy and the accuracy of the Lorentz
invariance are connected and that for a energy density $\rho_{vac.}=0$
Lorentz symmetry would be restored in the vacuum. In the cosmographic
model there is a similar connection. %
}. 

Nowadays the {}``vacuum'' has to be rather uniform on a cosmic scale.
There are many {}``vacuum''-states possible and one expects initially
quite an irregular structure of the {}``vacuum''. How can one understand
the observed homogeneity? 

First, the spreading out of condensation zones in the {}``vacuum''
could play an important role. From crystallization it is known that
condensation can bridge ten orders of magnitude in the laboratory
and the conditions for self-organizing might be orders of magnitude
better for the {}``vacuum'' state. 

Secondly there is an cosmological argument. A similar problem occurs
with the uniformity of the cosmic microwave background. To solve this
problem one introduced a scalar {}``inflaton'' field that  pushes
the universe initially into a rapid expansion (inflation). In this
way the required homogeneity in pre-inflation times becomes limited
in size. 

Actually the radiation emitted during condensation of the cosmographic
{}``vacuum'' might contribute to the expansion of the universe.
There are numerous cosmological aspects~%
\footnote{The repulsion of the extended Fermi structure covering the \char`\"{}vacuum\char`\"{}
might also play a role in the expansion of the universe~\cite{Linder:2008ya}.
A cosmological model with an interplay with a matter phase transition
and geometry was presented by Dreyer~\cite{Dreyer}.

Also the \char`\"{}vacuum\char`\"{} energy should somehow depend on
the gravitational potential blurring the distinction between Dark
Energy and Dark Matter~\cite{Alam:2008at}. An observation of a separation
of dark and visible matter in collisions of galactic clusters\cite{Bradac}
that contradicts a {}``modified Newtonian dynamics'' is no contradiction
to such a model in which the Dark Matter constitutes compressed Dark
Energy as the time scale of uncompressing the Dark Energy should should
be large (corresponding to the present tiny Dark Energy scale). %
} considered outside of the scope of the paper. 

To explain the known physics with its masses certain properties of
the {}``vacuum'' are required. But, as said, the knowledge about
the invisible, presumably strongly interacting {}``vacuum''-state
will stay rather limited. The detailed structure of the {}``vacuum''
will follow certain principles. How these principles could look like
will be discussed below.

\section{The Model for the Fermion Masses}

In the considered {}``cosmographic vacuum'' the Higgs fields $h{}_{ij}$
needed to obtain the fermion Dirac-masses is assumed to be dominated
by $f\bar{f}$ - condensates. It will be seen later that multi-fermion
states will actually have to have a comparable density. The assumption
is that these states are much less interacting with fermions as they
are more tidily bound. 

The condensate {}``vacuum'' does select a Lorentz system. The observed
effective Lorentz invariance of the visible world limits the interaction
with the {}``cosmographic vacuum'' at the available energies and
at the observable accuracy to a Lorentz scalar one. The huge geometrical
extension of vacuum states allows to use a scalar low energy effective
theory. Tensor couplings with non-zero rank can be ignored as they
involve derivatives that vanish in the limit $<(\sum r_{i})^{2}>_{\mathrm{condensate}}\rightarrow\infty$. 

To lowest order in perturbation theory there are the two contributions
available for the interaction with the condensate bound shown in figure
1 . The separation between visible world and {}``cosmographic vacuum''
is indicated by a box.

\begin{figure}
\begin{center}\includegraphics[%
  width=0.65\textwidth,
  keepaspectratio]{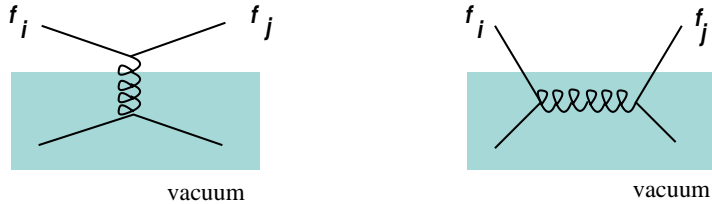}\end{center}

\caption{Fermion interactions with {}``cosmographic vacuum''}
\end{figure}

The first term involves tensors of rank 1 and vanishes in the considered
limit. The Fierz transformation of the second term includes a scalar
contribution that survives. The basic result is reassuring. As needed
the interaction with the {}``vacuum'' turns out to be flavor dependent,
with the mass of a fermion $m_{i}$ proportional to its fermion density
$\rho_{i}$ in the {}``cosmographic vacuum''. The exact form of
this dependence can be estimated in lowest order to be: $m_{i}\sim\rho_{i}g^{2}/\widetilde{M}^{2}$
where $\widetilde{M}$ is an effective mass of a boson exchanged in
the {}``vacuum'' and $g$ is the huge coupling of its scalar part~%
\footnote{Lorentz-scalar contributions can also originate in multiple gauge
boson contribution. In the considered GUT scheme the couplings are
known. Such processes stay at first in a perturbative regime and the
resulting hard Pomeron contribution is known to be small. %
}.

The condensate interaction interaction is strong and higher orders
have to be included. Presumably the cross section and the amplitude
depends on geometry. Relevant is the scale of the large mass of lightest
{}``techni-pion'' like bound state of the fermion {}``i'' and
anti-fermion {}``j'' directly coupling to the {}``vacuum'' state.
For the fermion masses its value is not well constrained as a small
geometrical size of a tidily bound state can be compensated by the
corresponding increase in density. In contrast to the energy density
the particle density can be huge. The Higgs density in the standard
model vacuum is also sizable.

Looking at the mass term more carefully there is an intricate interplay
between different scales. If a left-handed fermion would somehow transfer
from a zone with an empty vacuum where it is massless into a normal
{}``vacuum'' zone it would obtain a right-handed component. The
extra kinematic momentum slowing it down comes from pulling along
left-handed fermions and a right-handed anti-fermions in the {}``vacuum''
. A tiny non spin singlet component is excited in the {}``vacuum''.
To understand the situation we consider only one dimension. The bound
fermions move back and forth almost with the speed of light with alternating
handiness. The motion of the condensate states containing them changes
the times of the forward and backward motion and adjusts in this way
the left-handed right-handed balance in the laboratory system accordingly.
In this way transitions like $d{}_{R}\leftrightarrow d_{L}$ available
at the GUT scale in the condensate are transfered to the quark and
lepton scale.%
\begin{figure}
\begin{center}\cite{Porto}\includegraphics[%
  width=0.65\textwidth,
  keepaspectratio]{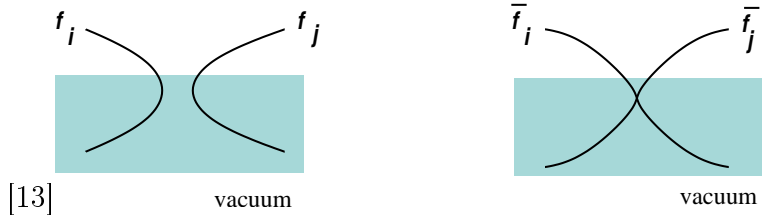}\end{center}

\caption{The equality of particle and antiparticle masses}
\end{figure}

In the considered zero-momentum limit there is no distinction between
incoming and outgoing particles as indicated in figure 2. Particles
and anti-particles have to have the same mass. \textit{CPT} is conserved
separately in the visible and invisible part of the world.

As said, the condensate {}``vacuum'' will have to be neutral and
colorless. The color structure can therefore be used to classify the
bound states of the condensate involving quarks as mesons, baryon
or more complicated objects. 

A similar argument exists for spins and angular momenta. As there
are no mass differences observed between left- and right-handed fermions
left and right components have to balance in the {}``vacuum'' and
the orbital angular momentum has to vanish. The {}``vacuum'' cannot
serve as spin reservoir and angular momentum is conserved separately
in the visible world. The {}``vacuum'' is not a source of parity
(P) invariance in the visible world. As this observation happens in
an arbitrary Lorentz system the equality has to hold separately for
any flavor. Baryonic states have to form scalar bosonic objects like
Cooper pairs. The observed situation seems natural. To minimize their
energy all {}``vacuum'' condensate boson has to be in the lowest
energy spin singlet state. 

However, the deposited and the picked up fermion do not have to be
identical. This is the source of a rich structure of the mass matrices.
Not all fermion index combinations $(i,j)$ can occur. The electric
and color neutrality of the condensate decomposes the mass matrix
into four separate sub-matrices: $M_{u,c,t}$, $M_{d,s,b}$, $M_{e,\mu,\tau}$,
and $M_{\nu(1),\nu(2),\nu(3)}$.

The result resembles now the standard approach. Diagonalization allows
to define flavor eigenstates conserved in neutral-current interactions
in the u\-su\-al way. Charge current interactions require to consider
mass matrices in a Cabibbo-rotated, non-diagonal basis leading to
flavor transitions in the usual way.

The diagonalization of the mass matrices was done after all orders
are summed. In this way flavor changing neutral currents are almost
absent. Nevertheless such currents can arise as the amplitudes responsible
for a mass matrix can depend on the virtuality of the fermion involved.
The mass-creating interaction is very localized and a miss-match in
virtuality with extended usual fermions causes rainbow like exchanges
that shield the mass in a somewhat flavor dependent way. However,
the gluon and the weak vector boson contribution also appears in the
standard model with its Higgs coupling matrices and is well-known
to be very tiny. A techni-pion like contribution could be a problem
as the mass of such $f\overline{f}$ Goldstone bosons~\cite{Leutwyler1996et}
would be significant and flavor dependent: \[
M_{\pi-like}^{2}\propto m_{\mathrm{Fermion}_{i,j}}\cdot B_{\mathrm{\mathrm{lowest}\, binding\, scale}}\]
As this mass is closer to the mass of typical momenta it could be
more effective introducing a flavor dependence than fermions. However,
as the techni-pion masses are here assumed to be considerably heavier
than the masses of the weak vector bosons their contributions should
be comparably tiny.

There are two fundamental differences to the standard approach. The
\textbf{\emph{first point}} concerns flavor conservation depicted
in figure 3. All flavors are conserved if the visible world and the
{}``vacuum'' are taken together. Only if the visible world is considered
separately \char`\"{}flavor''-changing processes appear. The {}``cosmographic
vacuum'' acts as reservoir. The initial flavor stability was obtained
during the diagonalization leading to the definition of the different
flavor states. The superscripts indicate that the states shown are
not the real charmed and up quark, but the states obtained in a charged
current interaction, i.e. the states of the transferred basis with
the $SU(2)$ partners of the $d$ respectively. $s$ quark. Now the
matrix element $M(c^{(s)}\to u^{(d)})$ is not balanced by the inverse
transition. Together with a charged current interaction the shown
change in the {}``vacuum'' is responsible  for the loss of strangeness.
\begin{floatingfigure}{0.5\columnwidth}%
\begin{center}\includegraphics[%
  bb=0bp 0bp 155bp 191bp,
  width=0.19\textwidth,
  keepaspectratio]{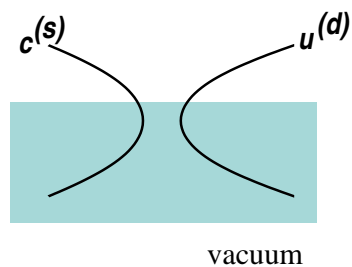}\end{center}

\caption{Flavor-changing contribution}\end{floatingfigure}%

As an example take an $s\overline{s}$ pair produced in $\ $$e^{+}e^{-}$
annihilation taking place in the visible world. Both strange quarks
will eventually annihilate from the {}``visible'' world ($K$ decays)
and leaves a corresponding pair in the {}``cosmographic vacuum'',
where the extra strange anti-strange content might eventually annihilate
as part of the cosmologically slow decay of the {}``cosmographic
vacuum''. 

Most novel is the \textbf{\emph{second point}}. It concerns {}``CP
violation''. Again CP violation arises as one restricts the consideration
to the visible world and ignores possible asymmetries in the {}``cosmographic
vacuum''. In the cosmographic model there is initially a particle--antiparticle
symmetry. The particle--antiparticle asymmetry in the visible world
in our zone of the universe is caused by the corresponding antiparticle--particle
asymmetry in the {}``vacuum''. CP violation is not cause for but
caused by the particle--antiparticle asymmetry. The situation is
subtle.

If in an asymmetric {}``vacuum'' the $\left<f_{i\,}\right>$ and
$\left<\overline{f_{i\,}}\right>$ contributions of the condensates
differ it can cause {}``CP violation'' in the visible world. The
point is illustrated in Figure 4. It lists the possible contributions
to the mass-matrix elements. As said above the {}``vacuum'' is parity
symmetric. Responsible for {}``CP violation'' is the charge asymmetric
difference between the first and the second line.

\begin{figure}
\begin{center}\includegraphics[%
  width=0.7\textwidth,
  height=40in,
  keepaspectratio]{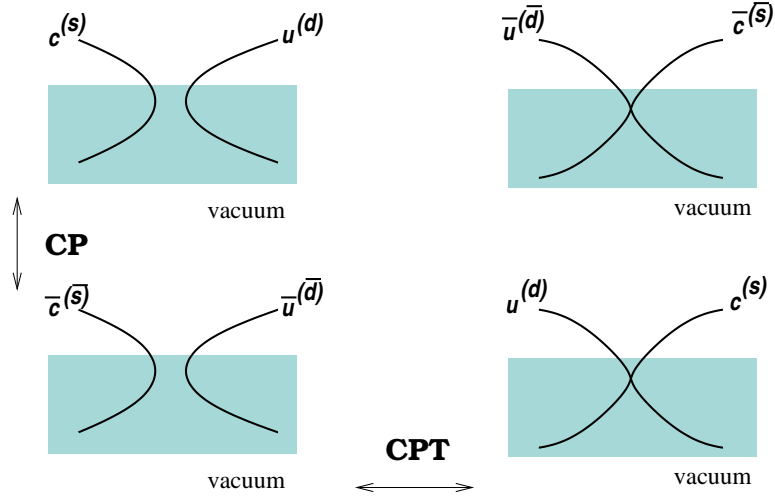}\end{center}

\caption{The symmetries of the flavor-changing contributions}
\end{figure}
Mirrowing the visible world there are more $\overline{d}$- than $d$-quarks
in the {}``vacuum'' condensate. For the {}``mesonic'' component
of the {}``vacuum'' which is assumed to dominate for fermion masses
flavor it means more $s\bar{d}$ than $\bar{s}d$. 

The specific evidence for CP violation comes from interference experiments.
Critical is the relative phase of the amplitudes $K_{short}^{0}\to\pi\pi$
and $K_{long}^{0}\to CP\, violation\to\pi\pi$ . To say it simple
in the $(K^{0},\overline{K^{0}})$ system a contribution with the
Pauli-matrix $\sigma_{x}$ introduces the CP eigenstates, a contribution
with the Pauli-matrix $\sigma_{z}=(\begin{array}{cc}
1\\
 & -1\end{array})$ introduces the CPT violation and a contribution with the Pauli-matrix
$\sigma_{y}=(\begin{array}{cc}
 & -i\\
i\end{array})$ introduces the CP violation. The data in the interference region
observe the phase and clearly show a CP-violating contribution%
\footnote{True CPT violation is presumably theoretical excluded. Possible is
a pseudo-CPT-violating term caused by a tiny matter-antimatter symmetric
part of the effective gravitational potential caused by the matter
surrounding us. However, a modern Eötvös experiment\cite{adelberger}
practically exclude such a $\sigma_{z}$~contribution.%
}.

Let us carefully consider the quantum mechanics involved. All mass
terms in the considered $3\cdot3$-matrices are a scalar, low momenta
limits of amplitudes. As there are no intermediate states the optical
theorem requires these amplitudes - taken by themselves - to be real.
As the {}``vacuum'' is involved, the {}``vacuum'' part of the
process has to be included in the consideration. As the {}``vacuum''
is a coherent state which is practically constant in space at the
considered scales there is no phase dependence on the locations where
strangeness is given respectively taken from the {}``vacuum'' . 

Interfering amplitudes obviously have the same initial and final states
and both $s_{\,}\overline{d}$ and $d_{\,}\overline{s}$ transition
appear equally often. Nevertheless there is a difference between \[
X_{S=0}\to K^{0}+\overline{K^{0}}\to K^{0}+K^{0}\to X_{S=0}+X_{S=0}\]
 with its C conjugate process. During the $\overline{K^{0}}\to K^{0}$
transition two $s$ quarks and two $\bar{d}$ quarks enter the {}``vacuum''
which are later compensated with two $\bar{s}$ quarks and two $d$
quarks during the decays of the two $K^{0}$ mesons. Important is
the different position in time. The phase difference is obtained by
the different intermediate state of the {}``vacuum''. As there are
more $\bar{d}$ in the {}``vacuum'' an additional $\bar{d}$ anti-quark
is energetically slightly less favorable than the corresponding $d$
quark. Without any new interaction a CP violating phase is obtained
corresponding on a qualitative level to the tiny experimental observation. 

For the total system including the {}``vacuum'' unitarity holds.
As the {}``vacuum'' has to change in a definite way the formal inclusion
of the {}``vacuum'' does not increase the dimension in the transition
matrices. With the usual argument one can show that the CP violation
in the $3\times3$ matrix in the quark sector involves only one parameter.
As in the usual description the unitarity triangle results which is
confirmed experimentally.

Predictions for the CP violation on the leptonic side are difficult.
Nothing is known about a lepton - antilepton asymmetry in the visible
or in the invisible world. The size of a microwave-like soft neutrino
background is unknown and the known atomic $e^{-}$ excess could be
compensated within the {}``visible world''. Anti-baryonic objects
(like Cooper-pairs $\bar{q}_{R}\bar{q}_{R}\bar{q}_{R}\bar{q}_{L}\bar{q}_{L}\bar{q}_{L}$)
could be the only available condensate states causing a purely hadronic
matter anti-matter asymmetry. The {}``vacuum'' states lives on a
GUT scale and it is also possible that a hadronic antiparticle preference
gets somehow transferred to the leptonic sector. 

\textbf{\emph{What could such a {}``vacuum'' look like?}} Fermions
obtain there masses from the interaction dominantly from the mesonic
{}``vacuum'' states. For the condensation of of very tightly bound
neutral, colorless $f_{i\,}\overline{f_{j}}$ pairs four ingredients
can be assumed: 

\begin{itemize}
\item If the standard SO$(10)\to$SU(5)
 scenario d-quark like bound states $D\bar{D}$ and leptonic bound
states $L\bar{L}$ require SO$(10)$ gauge boson while u-quark like
bound states $U\bar{U}$ can be formed just with SU(5) gauge bosons.
This might lead to a higher $U\bar{U}$ density explaining their higher
masses.
\item If in an initial random pre-vacuum process a fermion type somehow
happens to be favored, it obtains a heavier mass than the other fermions
which in turn allows a stronger condensation. This self-propelling
mass asymmetry could be a part of the explanation of the large variation
in the densities respectively masses. 
\item Like in crystals packaging geometry should play a role. Identical
fermion bound states could somehow be favored in a local crystal like
structure. 
\item The hierarchy of the observed masses\cite{Donoghue} will also reflect
fluctuations from the dominant $t\overline{t}$ condensate to other
fermion condensates. The probabilities of such fluctuations depend
on the overlap depending on the \char`\"{}similarity'' of the bound
states. Whether the binding is also supported by colors or identical
electric charges defines the \char`\"{}similarity''. In this way
the order of the $u$... , the $d...$, the $e...$and the $\nu...$fermion
masses might be understood . \\
This process will be more important for fermions with low densities.
As it tends to equalize the masses of different generations, it might
explain why the large mass ratios of $u,c,t$ quarks are not repeated
for the other fermions. 
\end{itemize}
But it is important to keep in mind that these are just general constraints
on a largely accidental cosmological {}``vacuum''. It is on the
same level as one can understand mountains in geography by considering
the motion of tectonic plates.

\section{The Vector-Boson Masses}

\begin{figure}
\begin{center}\includegraphics[%
  width=0.66\textwidth]{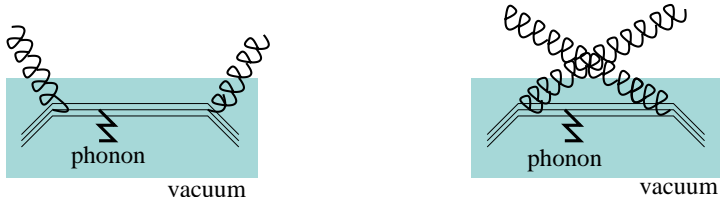}\end{center}

\caption{Vector-boson interactions with {}``cosmographic vacuum''}
\end{figure}
Vector bosons can also interact with fermions of the {}``cosmographic
vacuum'' in the way shown in figure 5. The neutrality of the {}``vacuum''
requires both vector bosons to have identical charges. As the weak
vector-boson masses are much smaller than the binding scale mixed
terms attaching to different fermions of the condensed state have
to be included. The mass creation is taken in analogy to light slowing
down passing through condensed matter. Phononic interactions within
the condensed matter are essential. The third component of the massive
vector boson is obtained in this way. They provide the $Q^{2}\to0$
pole needed in a formal theory to unprotect the vanishing vector masses. 

In the lowest-order in an operator product expansion one obtains a
term:\[
~\sum_{i,j}\left(\rho_{i,j}\,\vec{T}_{i}\,\vec{W}_{\mu}\,\,\vec{T}_{j}\vec{W}^{\mu}+\rho_{i,j}\, Q_{i}\, B_{\mu}\,\, Q_{i}\, B^{\mu}+\mathrm{mixed\, terms}\right)\]
where $\rho_{i,j}$ is the density of the fermions and where only
diagonal terms contribute. The lowest order contribution to the squared
vector-boson mass is then $M_{W}^{2}\sim\sum(\rho_{i,j}g^{2}/\widetilde{m})$,
where $g$ the effective coupling and $\widetilde{m}\sim0$ is the
effective mass of the bound state in the {}``vacuum'' . 

In spite of the apparent similarity the interactions responsible for
fermion- and vector-boson masses are quite different. Like a soft
hadronic interaction the exchange process responsible for fermion
masses depends of the geometrical extend of the target, while the
vector-boson interactions just count the charges of the condensed
{}``vacuum'' states %
\footnote{This meant as a rough approximation. It does not hold at scale when
the dipoles get resolved. P.e. the $U(1)_{X}$ boson could obtain
a mass from a right-handed neutrino condensate if the scales match
even so the condensate particle has no net $U(1)_{X}$ charge. %
}.

\begin{floatingfigure}{0.66\columnwidth}%
\begin{center}\includegraphics[%
  bb=0bp 0bp 699bp 150bp,
  width=0.6\textwidth,
  height=33mm]{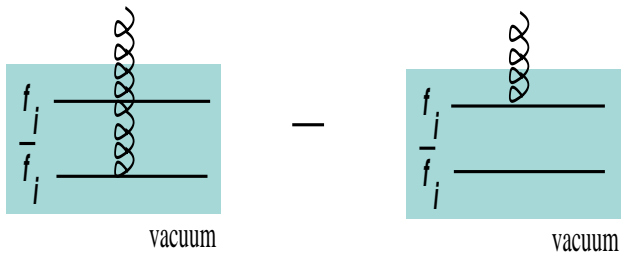}\end{center}

\caption{Mesonic contribution to the $B$ gauge field mass}

~\end{floatingfigure}%

Denoting $\# f_{L}$ the number of left-handed fermions i.e. $\# f_{L}=\# q_{L}+\# l_{L}$
and $\# f_{R}$ the number of right-handed fermions we define\begin{eqnarray*}
\Sigma f_{L} & = & \# f_{L}+\#\overline{f_{L}}\\
\Delta f_{L} & = & \# f_{L}-\#\overline{f_{L}}\\
\Delta f_{R} & = & \# f_{R}-\#\overline{f_{R}}\,\,.\end{eqnarray*}
The contributions to the vector boson masses are then:\[
M_{W}^{2}\sim(\frac{1}{2}\Sigma f_{L})^{2}\]
\[
M_{B}^{2}\sim\left(\frac{1}{3}\Delta q_{L}+\frac{1}{2}\Delta l_{L}+\frac{2}{3}\Delta u_{R}-\frac{1}{3}\Delta d{}_{R}-1\Delta l_{R}\right)^{2}\]
 where W is the SU(2) and $B$ is the U(1) gauge boson. 

The important point is a purely mesonic {}``vacuum'' yields $M_{B}=0$
as indicated in figure 6 . Baryonic contributions with non-vanishing
U(1) charge are needed. One candidate for such {}``vacuum'' states
are spinless neutron or antineutron Cooper pairs. 

Mixed $B$ -$W_{0}$ contributions terms exist in the masses matrix.
By diagonalization $M_{Z}^{2}=M_{W}^{2}+M_{B}^{2}$ and $M_{A}^{2}=0$
are obtained. The factorization form of these contributions yielding
$M_{A}^{2}=0$ is connected to the electric neutrality of the {}``vacuum''. 

The weak mixing angle parameterizes the relative size of the fermionic
contribution in the {}``vacuum''\[
\tan^{2}\theta_{{\rm {W}}}=M_{B}^{2}/M_{W}^{2}=\frac{<\left(\frac{1}{3}\Delta q_{L}+\frac{1}{2}\Delta l+\frac{2}{3}\Delta u_{R}-\frac{1}{3}\Delta d{}_{R}-1\Delta l_{R}\right)^{2}>}{<\frac{1}{4}\Sigma f_{L}^{2}>}\,\,.\]
The brackets mean averaging of the condensed states in the {}``vacuum''. 

The fact that the weak mixing angle $\tan^{2}\theta_{{\rm {W}}}=3/10+0.00071(14)$
is not tiny\cite{PDG} means that the baryonic and mesonic contributions
are comparable.

The weak mixing angle requires a large number of Cooper pairs of baryons
or anti-baryons. The net baryon number is arbitrary. It is appealing
to postulate a matter anti-matter symmetric world so that the imbalance
in the visible world is compensated by a corresponding imbalance in
the {}``vacuum''. In this way the non-vanishing weak mixing angle
requiring baryons or anti-baryons, the matter imbalance in the visible
world and CP invariance are connected.

\section{Experimental Expectations }

Can there be \textbf{Higgs-like} \textbf{particles} in \textbf{the
visible world? }

{\tiny ~}{\tiny \par}

In the {}``vacuum'' fluctuation in bosonic densities will be responsible
for the third component of the weak vector bosons. Three of such phononic
excitations in the condensate are needed. 

In principle such phononic excitations can be built with arbitrary
$f\bar{f}$ -pairs and there should be plenty of such techni-pion
like states with a mass vanishing at a GUT scale. Their precise mass
is difficult to estimate.

In analogy to the three special techni-pions associated with the vector-mesons
the mass of some of them might be in the $100\,$GeV range usually
assumed for the standard model Higgs. Most of them are presumably
much heavier: Expanding in small fermion masses one gets for such
pion-like states \cite{Leutwyler1996et}:\[
M_{\pi-like}^{2}\propto m_{\mathrm{fermion}}M_{\mathrm{lowest\, binding\, scale}}=10^{-3}\cdot10^{15}\mathrm{\,\, GeV^{2}}=(10^{6}\mathrm{GeV})^{2}\]
for MeV fermion masses and a GUT scale binding force. 

However, these estimates are very uncertain. If the neutrino mass
is $m_{\nu}\sim\Delta m_{\nu}=10^{-2}\mathrm{eV}$ the corresponding
lowest mass techni-pion would reach the 100 GeV range. Also $M_{\mathrm{lowest\, binding\, scale}}$
could be $M(U(1)_{X})$ instead of $M_{\mathrm{GUT}}$ which could
actually be much lower. It is not unlikely that the formula is not
really applicable. In contrast to the QCD scale entering for the usual
$\pi$ mass the corresponding $M_{\mathrm{lowest\, binding\, scale}}$
could have a flavor dependence. As in the vacuum the heavier flavors
could be more tidily bound and the mass ordering might even be inverted. 

It is probably not to difficult to distinguish these bosons from the
usual Higgs. They couple to the fermions in a completely distinct
way: The formation and the decay of quark-states typically involves
the same quark-types. They are private Higgs particles \cite{Porto}
essentially coupling only to one fermion type.

It is possible that the $\nu\bar{\nu}$ state has the lowest mass.
Formed from vector bosons the boson would predominantly decay into
neutrinos. Its signature would be that of an invisible Higgs \cite{ZhuS.-h}
with no constrain on its mass. If the $t\bar{t}$ state has the lowest
mass the signature would be a usual Higgs that cannot be found in
$b$-quark channels. The absence of abnormal backward scattering in
$e^{+}e^{-}$annihilation at LEP limits the corresponding leptonic
\char`\"{}Higgs''-boson to $M(H_{e^{+}e^{-}})>189\,{\rm {GeV}\,}$\cite{Bhabha}.
The large-transverse-momentum jet production at Fermilab limits $M(H_{\{ u\overline{u}\},\{ d\overline{u}\},\{ u\overline{d}\}\mathrm{or}\{ d\overline{d}\}})$
to an energy above $1{\rm \,{TeV}\,}$\cite{Fermilab}. At LHC a multi-TeV
range will be reached.

{\tiny ~}{\tiny \par}

Can \textbf{astrophysical observations be helpful}? Many cosmological
arguments are affected by the {}``cosmographic vacuum''. 

Except for uniformity in zones masses are accidental in the model.
How large are these zones? A valid test might be to look for changes
in the chiral symmetry breaking. Chiral symmetry breaking presumably
happened latest.

In our region of origin nuclear synthesis fixes $m_{e}/(m_{d}-m_{u})$
. Not excluded are synchronous changes expected in an evolving universe. 

The existing evidence for uniformity within our horizon is limited
as variations might not be seen. As variations in the chiral symmetry
breaking, that affect nucleon masses but not lepton masses, would
mix up nuclear synthesis such domains without observable stars would
not have been detected. Precision measurements might help to see still
admissible small variations in the nuclear radius%
\footnote{An appealing idea is to reanalyze the claimed $10^{-5}$ variation
in the fine structure constant \cite{Spectrum} as a zonal effect.
A {}``admissible'' $10^{-2}$ change in the nuclear mass could change
the hyperfine structure of the spectrum as $E_{hyperfine}\propto m_{e}/m_{p}E_{fine}$
in the right order of magnitude. %
}. 

With {}``dead'' zones in between anti-matter dominated parts of
the universe can no longer be excluded and experiments looking for
unusual antimatter components in very high energy cosmic rays could
be helpful.

\section{Conclusion }

The presented simple model is surprisingly successful in actual explanations
of many central particle physics observations. It would be nice to
understand the sketched condensation process in a more rigorous way
to not only pose but answer the question of the title. 

One purpose of this paper is to illustrate explicitely that mass generation
is wide open. If Higgs-like bosons are produced at LHC they might
decay in a completely unexpected way.

\end{document}